\documentclass[aps,pre,floatfix,superscriptaddress,showkeys,showpacs,twocolumn]{revtex4-1}
\pdfoutput=1 
\usepackage{graphicx}
\usepackage{hyperref}
\usepackage{amsfonts}
\usepackage{amssymb}
\usepackage{amsmath}

\newcommand{\mean}[1]{\langle #1\rangle}

\begin{document}

\title{Predicting epidemic outbreak from individual features of the spreaders}

\author{Renato Aparecido Pimentel da Silva}
\email[Corresponding author: ]{silva.renato@gmail.com}
\author{Matheus Palhares Viana}
\affiliation{Institute of Physics at S\~ao Carlos, University of S\~ao Paulo\\PO Box 369, S\~ao Carlos, S\~ao Paulo, 13560-970, Brazil}
\author{Luciano da Fontoura Costa}
\affiliation{Institute of Physics at S\~ao Carlos, University of S\~ao Paulo\\PO Box 369, S\~ao Carlos, S\~ao Paulo, 13560-970, Brazil}
\affiliation{National Institute of Science and Technology for Complex Systems, Rio de Janeiro, Brazil}

\date{\today}

\begin{abstract}
Knowing which individuals can be more efficient in spreading a pathogen throughout a determinate environment is a fundamental question in disease control. Indeed, over the last years the spread of epidemic diseases and its relationship with the topology of the involved system have been a recurrent topic in complex network theory, taking into account both network models and real-world data. In this paper we explore possible correlations between the heterogeneous spread of an epidemic disease governed by the susceptible-infected-recovered (SIR) model, and several attributes of the originating vertices, considering Erd\"os-R\'enyi (ER), Barab\'asi-Albert (BA) and random geometric graphs (RGG), as well as a real case of study, the US Air Transportation Network that comprises the US 500 busiest airports along with inter-connections. Initially, the heterogeneity of the spreading is achieved considering the RGG networks, in which we analytically derive an expression for the distribution of the spreading rates among the established contacts, by assuming that such rates decay exponentially with the distance that separates the individuals. Such distribution is also considered for the ER and BA models, where we observe topological effects on the correlations. In the case of the airport network, the spreading rates are empirically defined, assumed to be directly proportional to the seat availability. Among both the theoretical and the real networks considered, we observe a high correlation between the total epidemic prevalence and the degree, as well as the strength and the accessibility of the epidemic sources. For attributes such as the betweenness centrality and the $k$-shell index, however, the correlation depends on the topology considered.

\end{abstract}




\maketitle

\section{\label{sec:intro}Introduction}

Complex network theory has been in evidence over the last years, owing to its ability to explore discrete systems of interacting elements, considering a broad range of applications \cite{costa_et_al_2011_appli}. Particularly, it has proved to be a successful framework in the study of the relationship between topology and dynamics, a topic which has received growing attention. One example is to understand how individual characteristics and structural properties of networked systems influence on the spreading of diseases \cite{danon_2011}, a fundamental issue for disease control and eradication. For instance, it has been shown in the case of very large scale-free networks \cite{barabasi_1999} (at the thermodynamic limit) that the infection threshold vanishes \cite{pastor-satorras_2001}, meaning that the epidemic process persists no matter the magnitude of the infection rate. The eradication of epidemic processes in such systems should concentrate efforts on the {\it hubs}, i.e. the highly connected elements \cite{dezso_2002}. For small-world models -- which adequately portray social networks, where the understanding of epidemic processes is fundamental -- the problem has been analytically addressed by means of percolation theory \cite{moore_2000}. The authors demonstrated that an increase of the fraction of introduced shortcuts linking distant vertices progressively reduces both site and bond thresholds associated with disease spreading. Also related to ``small-world'' structures, M. Kitsak and his colleagues recently demonstrated that the capacity of a vertex to spread an epidemic disease is not necessarily a consequence of its degree or influence, given in terms of the betweenness centrality \cite{kitsak_2010}:  for certain real-world topologies, it was observed that the $k$-shell index \cite{carmi_2007} -- a degree-based measurement -- is capable to predict the best spreaders more accurately than both the previous attributes. The results so far mentioned focused exclusively on {\it homogeneous spreading}, when the transmission rate is the same across the whole system, disregarding eventual variations that may occur in terms of particular characteristics of elements and/or interactions. The heterogeneous case, however, has also received attention, and the studies in this case have either considered weighted networks \cite{gang_2005, schumm_2007, britton_2011} -- where the spreading rate across the connections is given by the weight of the connections itself or some function of the weight -- or by considering {\it meta-population} models, in which the hosts (individuals) are represented by means of interacting particles, which infect each other through reaction processes. Such particles are allowed to diffuse between different networked environments, promoting the spread of the disease in the space. Examples of the meta-population approach include the study of the propagation of the foot-and-mouth disease among the cattle and then across different farms in the United Kingdom, due to livestock movements \cite{kao_2006}, and the Ref. \cite{colizza_2007}, where the diffusive spreading in both scale-free networks and the US Air Transportation Network are investigated. Regarding the use of weighted networks, on the other hand, it has been found that on scale-free networks where the distribution of the infection rates across the connections obeys a power-law, the spreading is slower when compared to the homogeneous counterpart \cite{gang_2005}. The spreading is also hierarchical in terms of the strength -- the {\it weighted degree}, i.e. the sum of the weights of attached connections  -- of the vertices: those with higher strengths are infected at first, whereas vertices with lower strengths are infected only at later stages of the epidemics.  In a recent result \cite{britton_2011}, a weighted configuration model (e.g. \cite{molloy_1995}) is introduced and an expression for the corresponding epidemic threshold is derived. 

In this paper we address the correlation between several individual attributes and the spreading potential of network vertices, considering SIR (susceptible-infected-recovered) dynamics \cite{diekmann_2000, *keeling_2008} and the heterogeneous assumption. In the SIR model, infected individuals are allowed to transmit the disease to connected susceptible counterparts at rate $\beta$, and spontaneously recover at rate $\mu$. $\beta$ and $\mu$ correspond to the control parameters of the model, being denominated {\it transmission rate} and {\it recovery} (or {\it removal}) {\it rate}, respectively. The heterogeneous spreading hypothesis here adopted assumes that the transmission rate $\beta_{ij}$ of an infected individual $i$ to transmit the pathogen to a susceptible contact $j$ is not the same for all established connections $(i,j)$. Indeed, a realistic approach for such dynamics should consider heterogeneity, in an attempt to accommodate aspects such as different levels of interaction among the elements, as well as healthy conditions at the individual level. In this text we deal with undirected networks, meaning that $\beta_{ij}=\beta_{ji}, \forall (i,j)$, and discard differences occurring at individual level, employing the same recovery rate $\mu$ for all the vertices. In order to emulate the heterogeneity, a geographic network model is considered, where the transmission rate decays exponentially with the distance separating interconnected vertices -- in other words, the connections define the existence of contact between the individuals, and the length of the connections are employed to determinate  the transmission rate. An analytic expression for the transmission rate distribution is derived, and subsequently the heterogeneous spreading considering the same distribution for the transmission rates  is also studied in non-geographic models, namely Barab\'asi-Albert scale-free networks and Erd\"os-R\'enyi random graphs. Furthermore, a real network is also considered, the US Air Transportation Network defined by the 500 busiest American airports, interconnected by weighted connections defining the yearly seat availability among every two locations \cite{colizza_2007}. In this case, the distribution of the transmission rates is given in terms of the seat offer. The spreading potential of each vertex is quantified in terms of the total prevalence of the epidemic process, i.e. the fraction of vertices which have been infected during the epidemic outbreak. Six vertex attributes are considered, namely the already mentioned degree, $k$-shell index and betweenness centrality, plus the weighted degree or strength, the clustering coefficient (the density of connections at local level) and the recently proposed measure of accessibility \cite{travencolo_2008}, which estimates the number of individuals effectively reached by paths of a determinate length, departing from the vertex. High correlation is verified between the epidemic prevalence and half the measurements, considering all the network models and the US Air Transportation Network: the degree, the strength and the accessibility. On the other hand, the inter-connectivity at local/regional level, expressed by the clustering coefficient, little informs about the spreading potential of the individuals, since low correlation with epidemic prevalence is observed in all the cases. The prediction ability of betweenness centrality is reduced in the case of geographic networks, where distant vertices do not make contact. In contrast to what has been observed for real small-world networks \cite{kitsak_2010}, the $k$-shell index can not be considered in the case of scale-free model, where all the vertices feature the same value of this measure, even presenting distinct behavior as epidemic spreaders. Since the correlation between an individual aspect and the dynamic behavior of the spreading process depends on the system topology, our analysis suggests one should avoid considering a single aspect when predicting the potential spreaders.

Our article is organized as follows: the next section describes the complex network models used in the text, focusing on the geographic network model, in which we derive the analytic expression for the distribution of the transmission rates across the connections. The characterization of vertices by means of measurements is also addressed. Afterwards, we describe the SIR model and how it is applied at such networks, considering the epidemic thresholds observed. Finally, the results are presented and discussed, for the generated networks so far described as well as for the airport network.

\section{\label{sec:net}Networks: theoretical modeling and characterization}

In order to vary the transmission rate across different pairs of inter-connected vertices, we first consider a spatial or geographic model, where such rate decreases with the distance between the vertices. The approach is extended to two widely-known ``non-spatial'' network models, Erd\"os-R\'enyi random graphs and Barab\'asi-Albert scale-free networks. The description of the theoretical models is given below.

\subsection{\label{sec:model}Network models}

Epidemics spread typically in space-embedded systems, where the {\it proximity} or even the {\it physical contact} between individuals is fundamental on the transmission of pathogens. Examples include respiratory diseases, such as influenza; plant diseases, e.g. citrus greening disease -- bacterial plant disease mainly spread via psyllid insects \cite{halbert_2004}; etc. In this paper adopt the random geometric graph (RGG) \cite{dall_2002, *barthelemy_2011}.

In the RGG model, $N$ vertices are distributed at random coordinates ${\bf x}_i = (x_{i1},x_{i2},\dots,x_{id})^t,\,x_{ij} \in [0,1)$ inside a $d$-dimensional unit cube. Links are established between every pair of vertices whenever the Euclidean distance separating both the individuals is not greater than a predefined threshold $R$. Note that the $d$-dimensional hyper-volume of the unit cube is 1, so that the probability of a vertex $j$ be distant at most $d_r$ from another vertex $i$ is equal to the hyper-volume $V_{dr}$ of the $d$-dimensional hyper-sphere with radius $d_r$ centered at ${\bf x}_i$. In this paper we deal with $d=2$, then $P(0\leq |{\bf x}_i-{\bf x}_j| \leq d_r) = \pi d_r{}^2$. Therefore, two vertices $i$ and $j$ picked at random are connected with probability

\begin{equation}\label{eq:rgg}
p_R = \pi R{}^2 \equiv P(0\leq |{\bf x}_i-{\bf x}_j| \leq R)\,.
\end{equation}

If the average number of connections per vertex is $\langle k\rangle_{RGG}$, then we have $p_R =  \langle k\rangle_{RGG}/(N-1)$\footnote{If edges are defined with probability $p$, then on a graph with $N$ vertices the expected number of edges is $pN(N-1)/2$. In other words, the average degree of the graph is $p(N-1)$.}, which allows the choice of the cutoff distance $R$ in terms of connectivity and the network size $N$: 

\[R\left(\langle k\rangle_{RGG},N\right)=\sqrt{\frac{\langle k\rangle_{RGG}}{\pi(N-1)}}\,.\]

Aspects such as the proximity or the amount of contact between individuals are crucial determinants for epidemic spreading in networked systems. Therefore, it is expected that increasing distance reduces or to a large extent eliminates the chance of an infected individual to transmit the pathogen to a susceptible counterpart. Indeed, when modeling the spread of disease, a straightforward approach is to consider the {\it transmission rate} $\beta_{ij}$ between two individuals $i$ and $j$ as being a decreasing function of the distance $|{\bf x}_i-{\bf x}_j|$. In the current paper, when considering the RGG model, the decay of such rate is given by the exponential function

\begin{equation}\label{eq:weight}
\beta_{ij} = \exp{\left(-A\frac{|{\bf x}_i-{\bf x}_j|}{R}\right)}\,,
\end{equation}

\noindent where $A$ is a positive constant that controls the global average probability $\langle\beta_{ij}\rangle$ and $R$ is the cutoff distance as previously described. Since the {\it length} $L_{ij}\equiv|{\bf x}_i-{\bf x}_j|$ of each edge in the RGG model is such that $0\leq L_{ij}\leq R$, we have $\exp(-A)\leq\beta_{ij}\leq 1$. It is possible to derive the distribution $\beta_{ij}$ for all the connections present in the 2-dimensional RGG graph: from Eq. (\ref{eq:rgg}) we have, for the edge length $L_{ij}$,
\[
P(0\leq L_{ij}\leq L) = \left(\frac{L}{R}\right)^2\,,
\]

\noindent with $E(L_{ij})=\langle L_{ij}\rangle=2R/3$. By using Eq. (\ref{eq:weight}) -- note that $\beta_{ij}=\exp{(-AL_{ij}/R)}$ -- we obtain $(L/R)^2 = P(\exp{(-AL/R) \leq \beta_{ij} \leq 1)}$ which yields, defining $\beta = \exp(-AL/R)$,

\begin{eqnarray*}
P(\beta\leq \beta_{ij} \leq 1) &= &\frac{1}{A^2}\ln^2{\left(\frac{1}{\beta}\right)}\\
&= & \frac{1}{A^2}\ln^2{\beta}\,,
\end{eqnarray*}

\noindent $\beta\geq \exp(-A)$, which yields to

\begin{equation}\label{eq:cdf}
F(\beta)=P\left(\exp(-A) \leq \beta_{ij} \leq \beta\right) = 1-\frac{1}{A^2}\ln^2{\beta}\,. 
\end{equation}

From the CDF above the value $\langle \beta_{ij}\rangle$ can be derived, and it is given by

\begin{equation*}
\langle \beta_{ij}\rangle = \frac{2}{A^2}\left[1-(1+A)\exp(-A)\right]\,.
\end{equation*}

Therefore, one can generate a RGG ensemble with a given average transmission rate $\langle \beta_{ij}\rangle$  by choosing the appropriate positive value of $A$ that solves the expression above, provided $\beta_{ij}$ is given by Eq. (\ref{eq:weight}) for every $i, j$. 

\begin{figure}
\centering
\includegraphics[width=\columnwidth]{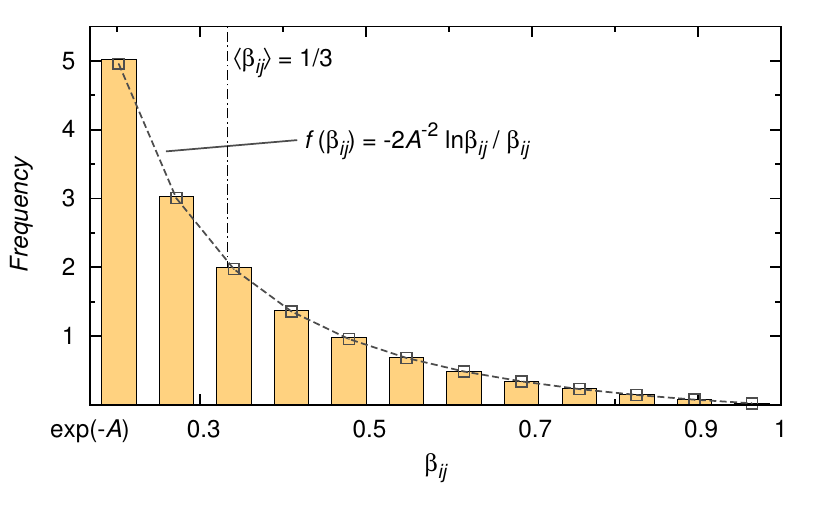}
\caption{\label{fig:beta-dist}The distribution of the transmission rate $\beta_{ij}$, as defined in the text for RGG and extended for ER and BA networks. Vertical bars give the observed frequency, considering all the RGG networks sampled for the current study, and the dashed line corresponds to the derivative of the CDF given by Eq. (\ref{eq:cdf}).}

\end{figure}

In the text we discuss the case $\langle\beta_{ij}\rangle = 1/3$, so ${A\approx 1.79055}$. Fig. \ref{fig:beta-dist} depicts the histogram of $\beta_{ij}$, considering all the RGG network samples used in the text, as well as the corresponding analytic expression, given by the derivative of Eq. (\ref{eq:cdf}).

\subsubsection*{Extension to scale-free network and random graphs}

In order to observe the topological aspects of the networked system as a whole on the investigated correlations, we also consider two widely-known ``non-geographic'' models, in addition to the geographic model previously presented: Erd\"os-R\'enyi random graphs (ER) \cite{erdos_1959} and Barab\'asi-Albert scale-free networks (BA) \cite{barabasi_1999}. In the ER model, one starts with $N$ vertices, and for each pair of vertex an edge is added with probability $p$, so that at the end $pN(N-1)/2$ links are established and the {\it average degree} $\langle k\rangle$ of the network, that is, the average number of connections per vertex, is $\langle k\rangle_{ER}=p(N-1)$; BA model is characterized by a growth-process where departing from $m_0$ vertices, a new vertex is added at each step, being linked to $m\leq m_0$ vertices. This attachment gives preference to highly connected ones, i.e. the probability of the new vertex $N$ to be connected to an existent vertex $i$ is $p_{N,i}=k(i)/\sum_{j=1}^{N-1} k(j)$, where $k(i)$ denotes the number of connections previously attached to $i$. After $t$ steps the network has $N=m_0+t$ vertices and $mt$ edges. For increasing values of $N$, $\langle k\rangle_{BA}$ tends to $2m$, and the ``signature'' of the scale-freedom emerges: the distribution of vertex degrees follows a power law -- $P(k) \sim k^{-\gamma}$, with $\gamma=3$ -- a consequence of preferential attachment policy. The skewed distribution diverges from that found on ER networks, a Poisson distribution peaked at $\langle k\rangle_{ER}$. In other words, while in BA networks the vertex degree $k$ may span across several orders of scale, for an ER graph it remains at the same order of $\langle k\rangle_{ER}$. The heterogeneous spreading is extended for ER random graphs and BA scale-free networks by assuming the transmission rate $\beta_{ij}$ as distributed according to Eq. (\ref{eq:cdf}). Specifically, we at first generate a network according to one model, then we assign for each link $(i,j)$ a randomly chosen weight (the 
spreading rate) based on the cumulative distribution function (\ref{eq:cdf}).

\subsection{\label{sec:acc}Individual attributes: vertex measurements}

The topology of complex networks, as well as particular aspects of the vertices and interactions that define such systems, is better understood by means of its characterization with the use of quantitative measurements \cite{costa_et_al_2007_meas}. As mentioned earlier, here we observe the prediction capabilities of six vertex measurements:

\begin{itemize}
 \item Vertex degree ($k$): the number of connections attached to the vertex. In other words, the number of immediate or nearest neighbors.
 \item Strength ($s$): the ``weighted degree'', defined by the total weight of connections attached to the vertex: $s(i)=\sum_j w_{ij}$. As in the text the weights correspond to the transmission rates among contacts, here we have $s(i)=\sum_j\beta_{ij}$.
 \item Clustering coefficient ($C$): provides the density of connections among the nearest neighbors of a vertex $i$, and $i$ itself. $C(i)= 2\ell(i)/[k(i)(k(i)-1)]$, where $\ell(i)\leq [k(i)(k(i)-1)]/2$ is the amount of such connections. 
 \item $k$-shell index ($k_S$): degree-based measurement calculated through the $k$-shell decomposition. If $m$ is the minimum degree of the network ($k(i)\geq m,\forall i$), then all vertices $j$ such that $k(j)=m$ are given $k$-shell index $k_S(j)=m$, and subsequently removed from the network. After removal, it is possible that a new set of vertices with degree $m$ emerges. Vertices belonging to such set are then given index $k_S=m$ and removed, and the process is repeated until no more vertices with $m$ connections are observed. The next step of decomposition involves vertices with degree $m+1$, and so on, until all the vertices are attributed a $k$-shell index. Note that $k_S(i)\leq k(i)$. Such decomposition define layers along the network, each one  associated to a corresponding $k_S$ value. Layers with higher $k_S$ are the innermost, defining a ``core'' on the network. More details in \cite{carmi_2007, kitsak_2010}.
\item Betweenness centrality ($C_B$): For a vertex $i$, the ratio of shortest paths connecting two vertices $j$ and $k$ that passes through $i$ to the total number of shortest paths linking $j$ and $k$, averaged over all pairs $j$, $k$ \cite{freeman_1977}.
\item Accessibility ($A_H$): the accessibility of a vertex $i$, as the name suggests, is an estimate of how many vertices can be ``accessed'' through random walks departing from $i$ after $H$ steps, and it is given by \cite{travencolo_2008}

\[
 A_H(i) = \exp{\left(-\sum_{j=1}^N P_H(i,j)\log{P_H(i,j)}\right)}\,,
\]

\noindent where $\sum_j P_H(i,j)=1$. $P_H(i,j)$ is the probability that a particular vertex $j$ is reached after $H$ steps by an agent whose walk starts at $i$, i.e. passing through $H-1$ intermediate vertices. In the case of weighted graphs, such random walks can be preferential, in the sense that if at a given step the agent is at vertex $i$, then it will be located at a nearest-neighbor $q$ of $i$ with probability proportional to the weight of the connection $(i,q)$. The term inside the exponential corresponds to the entropy of the probabilities $P_H(i,j)$ and it is denominated the {\it diversity} of $i$. Under the hypothesis that, departing from $i$, all the $N-1$ remaining vertices of the network are reachable after $H$ steps with the same probability, then the diversity will assume its maximum value, $-\log(1/(N-1))\equiv log(N-1)$, and hence the accessibility, $A_H(i)=N-1$. On the other hand, if in $H$ steps the agent can reach only the same vertex, then the accessibility of $i$ is minimum and $A_H(i)=1$. If $D$ is the diameter of the network -- that is, the maximum length of the shortest paths between any pairs of vertices. Note that several measures of accessibility can be estimated for a single vertex, considering $H=1,2,\dots$. For this paper we fix the value of $H$ as the average shortest path length of the network.

\end{itemize}

\section{SIR disease spreading}


In this text we adopt the SIR (susceptible-infected-recovered) model \cite{keeling_2008} for the simulation of spreading processes over the networks. The population is divided into three classes or {\it compartments}, the {\it susceptible}, {\it infected} and {\it recovered} individuals. A susceptible individual is a healthy element which is allowed to contract the disease. An infected individual, on the other hand, is a contaminated element capable of transmit the disease to its susceptible contacts. In this paper, the contagion routes are given by the connections $(i,j)$ defined in the networks, such that an infected individual $i$ transmits the pathogen to a susceptible contact $j$ at a rate $\beta_{ij}$. Infected individuals recover from the disease at rate $\mu$, becoming immune to further infections, and thus are eliminated from the epidemic process. If $\mu> 0$, the epidemic process always terminates, i.e. the {\it prevalence} of infected individuals $i(t)$ always becomes null for sufficiently large $t$. As the goal is to observe the impact of the epidemic process on the network, in terms of characteristics of the epidemic sources, here the epidemic process is promoted by infecting a single vertex, thus $i(0)=1/N\approx 0,\,s(0)\approx 1,\,r(0)=0$. $s(t)$ and $r(t)$ are the density of susceptible and recovered elements at the time $t$, respectively. Here we apply the discrete approach \cite{danon_2011} to simulate the epidemic process: At time $t$, an infected vertex $i$ transmits the disease to each of its susceptible contacts $j$ with probability $\beta_{ij}dt$, where $dt\ll 1$ is the time-step of the computational process. Such event is considered for all contact-links $(i,j)$ that connect infected and susceptible individuals. At the same time, infected individuals are allowed to spontaneously become immune (recovered) with probability $\mu dt$. The contamination and recovery processes are then repeated at the time $t+dt$ and so on, until no infected individuals are present in the network. The damage of the epidemics to the system is then quantified in terms of the {\it total epidemic prevalence} or {\it epidemic outbreak size} $r_{\infty}=\lim_{t\rightarrow\infty}r(t)$, which corresponds to the totality of individuals contaminated until the end of the epidemic process.

As seen in the previous section, here we assume that the transmission rate $\beta_{ij}$ varies across the connections of the networks. In the {\it homogeneous} assumption, i.e. when $\beta_{ij}\equiv \beta, \forall i,j$, several aspects of SIR model in complex networks are known, including the existence of an {\it epidemic threshold}, a value $\lambda_c$ such that the infection only spreads across the network if $\beta/\mu\geq\lambda_c$ -- for $\beta/\mu<\lambda_c$, on the other hand, no endemic phase occurs. For networks with no degree correlations, at mean-field level the epidemic threshold is given by \cite{barrat_2008}

\begin{equation}\label{eq:thresh}
\lambda_c = \frac{\langle k\rangle}{\langle k^2\rangle-\langle k\rangle}\,.
\end{equation}

\begin{figure}
\centering
\includegraphics[width=\columnwidth]{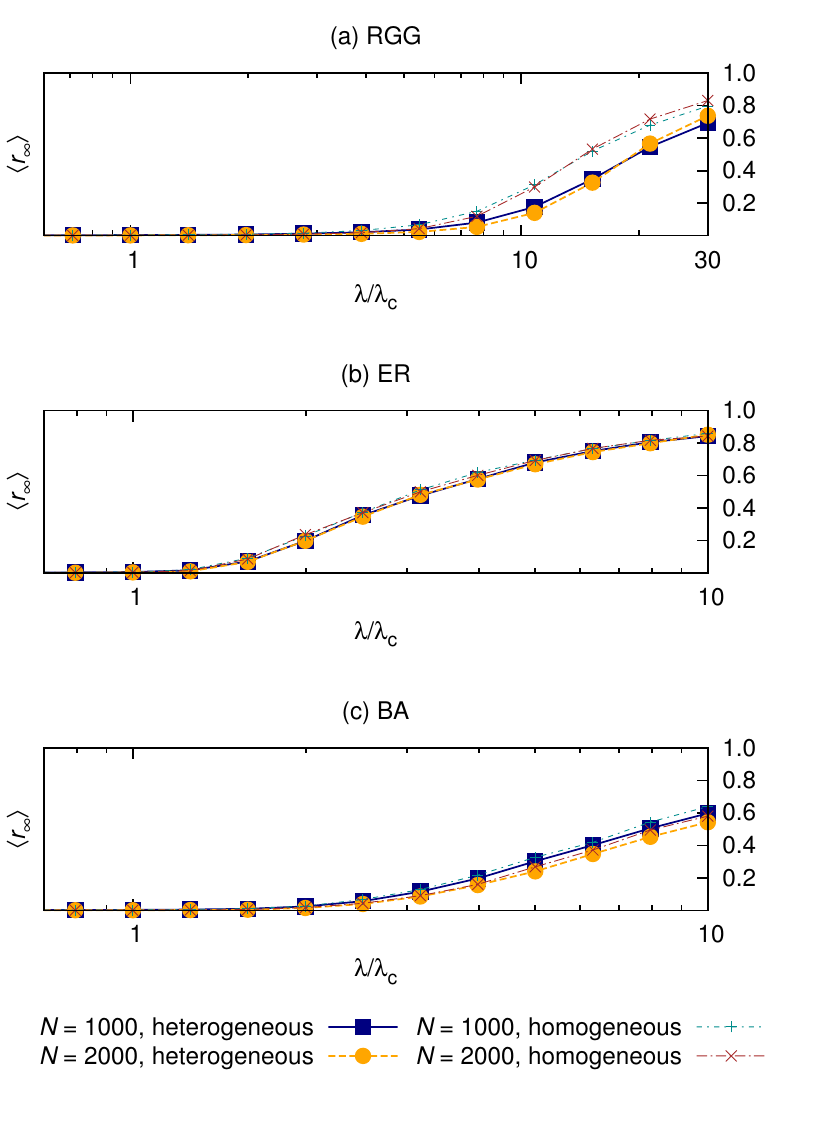}
\caption{\label{fig:thresh}Average epidemic outbreak for different $\lambda={\beta}/{\mu}$, for the heterogeneous spreading processes considered in the text (varying transmission rate $\beta_{ij}$ with $\langle\beta_{ij}\rangle=1/3$) and for homogeneous spreading considering $\beta_{ij}=\beta=1/3,\forall i,j$. The threshold value $\lambda_c$ is estimated as given on Eq. (\ref{eq:thresh}).}
\end{figure}

\noindent The existence of an epidemic threshold is also expected in the heterogeneous assumption here considered, since for the networks here explored all the rates of infection $\beta_{ij}$ lie between 0 and 1 (see previous section). Indeed, for ER and BA networks, we observed that the estimate given by Eq. (\ref{eq:thresh}) is adequate for both the cases: for the homogeneous infection, we set $\beta = \langle\beta_{ij}\rangle = 1/3$, and varied the fraction $\lambda=\beta/\mu$ in the interval $1/10\lambda_c,\dots,10\lambda_c$ by changing the recovery rate $\mu$. The same values of $\mu$ were used for the heterogeneous case. We observed that in the case of ER and BA networks, the average epidemic outbreak size was the same for both the homogeneous and heterogeneous cases. For the geographic networks generated by RGG model, on the other hand, the heterogeneity of the spread along the contacts increased the resilience of these systems to attack, since higher values $\lambda$ are required for epidemic spread. Also, RGG are more resilient than ER and BA graphs. For each network, several ensembles were performed at each value $\lambda$, considering random source vertices, and the analysis is illustrated in Fig. \ref{fig:thresh}.  In this text the epidemic behavior is explored right above the threshold, where the average outbreak is small. At such condition, the epidemic behavior is highly influenced by properties of the source vertex, in opposition of what is observed for higher values of $\lambda$, when the epidemic spread is always catastrophic for the system.

\begin{table*}[!ht]

\caption{\label{tab:net}Characterization of network models. 30 samples each class are used in the text. Average values, from the left to the right: the number of vertices $N$, the number of links $\ell$, the average degree $\langle k\rangle$, the average squared degree $\langle k^2\rangle$, the average shortest path length $L$, the epidemic threshold $\lambda_c$ -- as estimated by Eq. (\ref{eq:thresh})) -- and the value $\lambda$ employed on the simulations, in terms of $\lambda_c$.}
\begin{ruledtabular} 
\begin{tabular}{cccccccc}
model & $N$ & $\ell$ & $\langle k\rangle$ & $\langle k^2\rangle$ & $L$ & $\lambda_c$ & $\lambda/\lambda_c$ \\
\hline 
RGG, 1000 vertices & 985.1 $\pm$ 4.216 & 2867 $\pm$ 53.60 & 5.82 $\pm$ 0.108 & 39.62 $\pm$ 1.745 & 22.40 $\pm$ 2.287 & 0.172 $\pm$ 0.005 & 5\\
RGG, 2000 vertices & 1969 $\pm$ 6.184 & 5827 $\pm$ 79.24 & 5.920 $\pm$ 0.081 & 40.99 $\pm$ 1.345 & 29.91 $\pm$ 2.429 & 0.169 $\pm$ 0.004 & 5\\
ER, 1000 vertices & 997.4 $\pm$ 1.569 & 2997 $\pm$ 1.762 & 6.009 $\pm$ 0.010 & 42.02 $\pm$ 0.246 & 4.056 $\pm$ 0.007 & 0.167 $\pm$ 0.001 & 1.25\\
ER, 2000 vertices & 1995 $\pm$ 2.024 & 5997 $\pm$ 2.280 & 6.011 $\pm$ 0.007 & 42.05 $\pm$ 0.219 & 4.444 $\pm$ 0.007 & 0.167 $\pm$ 0.001 & 1.25\\
BA, 1000 vertices & 1000 & 2991 & 5.982 & 84.51 $\pm$ 22.36 & 3.492 $\pm$ 0.042 & 0.077 $\pm$ 0.006 & 2\\
BA, 2000 vertices & 2000 & 5991 & 5.991 & 92.47 $\pm$ 3.897 & 3.736 $\pm$ 0.022 & 0.069 $\pm$ 0.003 & 2
\end{tabular}
\end{ruledtabular}
\end{table*}

\section{Results and discussion}

In this text the behavior of epidemic spread has been observed for the network models previously described in Sec. \ref{sec:net}, where 30 samples were considered each model, for system sizes $N=1000$ and $N=2000$. In the case of ER graphs and RGG, only the largest connected component was taken into account -- we considered ER and RGG samples whose largest connect component size is at least 98\% of the total size $N$. Some aspects of such networks are highlighted in Table \ref{tab:net}, where the epidemic threshold for homogeneous spreading and random networks is also given, from Eq. (\ref{eq:thresh}).

\begin{figure}[!ht]
\centering
\includegraphics[width=\columnwidth]{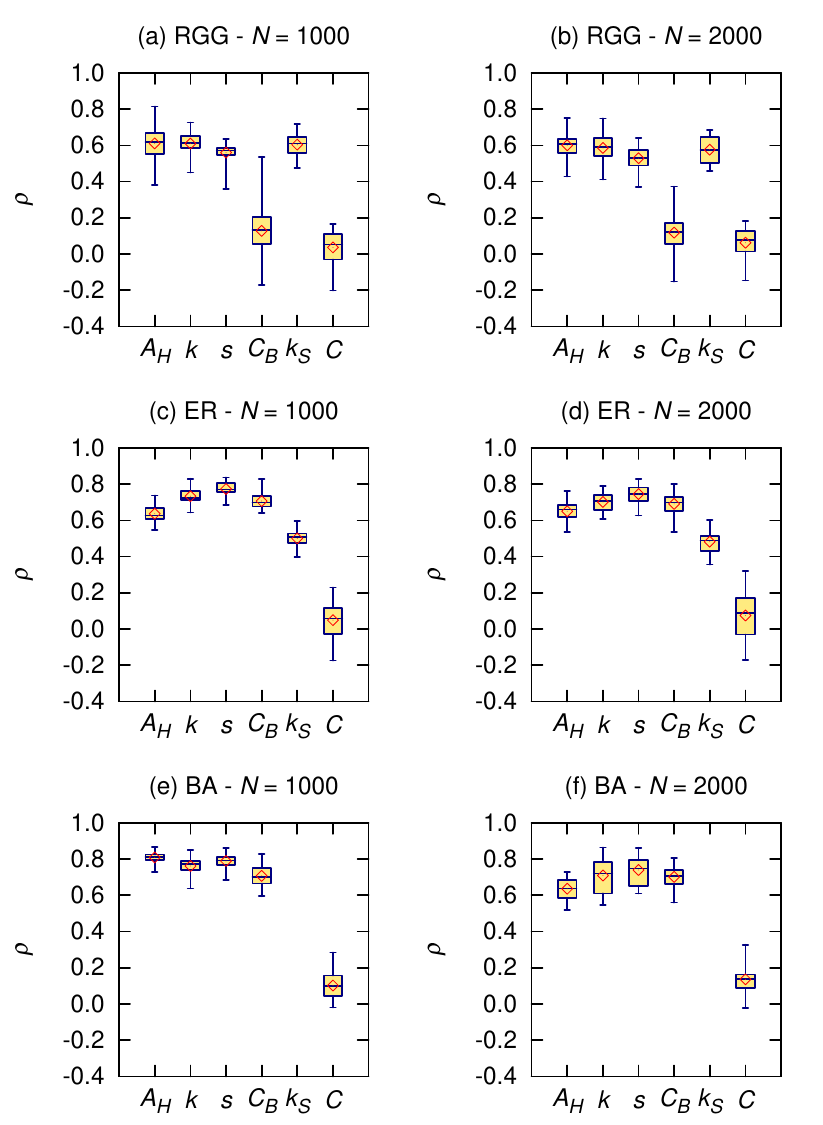}
\caption{\label{fig:whiskers}Behavior of Pearson correlation coefficients $\rho$ between the epidemic total prevalence and the accessibility $A_H$, the degree $k$, the strength $s$, the betweenness centrality $C_B$, the $k$-shell index $k_S$ and the clustering coefficient $C$, across all the samples for each network class. }
\end{figure}

\begin{figure}[!ht]
\centering
\includegraphics[width=\columnwidth]{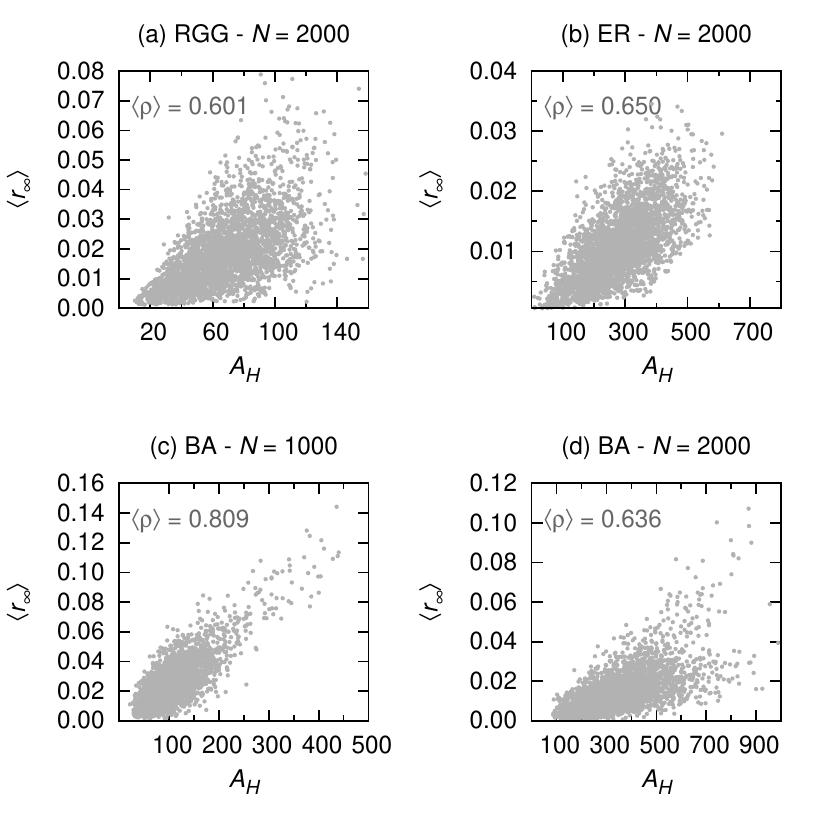}
\caption{\label{fig:acc}Average epidemic prevalence $\mean{r_\infty(i)}$ as function of vertex accessibility $A_H(i)$, overall observed values. Results for network size $N=1000$ in the case of ER and RGG graphs omitted for clarity, due to results such structures showed, similar to the $N=2000$ case. The Pearson coefficient $\rho({\bf r_\infty},{\bf A_H})$, averaged over all the samples for each case, is given.}
\end{figure}

A subset of 100 vertices were randomly chosen for each network sample, and subsequently the spreading potential of each selected vertex $i$ was estimated in terms of the average epidemic prevalence $\mean{r_\infty(i)}$ observed on epidemic spreads having $i$ as {\it source} -- i.e. started at $i$. We traced the relationship between the prevalence, considering all the selected vertices, and individual attributes of the latter, by means of the Pearson correlation coefficient $\rho({\bf r_\infty},{\bf x})$, where ${\bf r_\infty}=\{\mean{r_\infty(i_1)},\,\mean{r_\infty(i_2)},\dots,\,\mean{r_\infty(i_{100})}\}$ and ${\bf x}=\{x(i_1),x(i_2),\dots,\,x(i_{100})\}$. $x(i)$ is either one of the six vertex attributes considered, e.g. the degree $k$, the strength $s$ and so on. The analysis of correlation is depicted in Fig. \ref{fig:whiskers}, where the distributions of such correlation coefficients across all the samples for each network model are illustrated in terms of box-and-whisker plots. It is clear that the local inter-connectivity -- estimated by means of clustering coefficient $C$ -- tells little about the spreading potential of each source vertex, as the value $\rho({\bf r_\infty},{\bf C})$  low regardless of the network topology. The behavior of other vertex attributes varies according to the considered network topology. This is particularly evident in the case of the betweenness centrality $C_B$, which features significantly lower correlation for RGG networks, in comparison to ER and BA structures, possibly an effect of the spatial limitations imposed on the connectivity. The $k$-shell index proved inadequate for such analysis in the case of BA networks, where all the vertices featured the same $k$-shell index, $k_S=3$. The Pearson correlation coefficient  $\rho({\bf r_\infty},{\bf k_S})$ is undefined in this case, explaining the absence of boxes and whiskers for $k_S$ in Figs. \ref{fig:whiskers}(e)--(f). The $k$-shell index also features lower correlation when compared to other measures for ER random graphs. In terms of network size $N$, no significant differences in the results are observed for ER and RGG networks, in opposition to BA networks, which suggest size effects for this case. Indeed, as the size of BA network increases, the average squared degree $\mean{k^2}$ tends to diverge, which ultimately yields to the null epidemic threshold theoretically derived at the thermodynamic limit -- a decrease in $\lambda_c$ is observed  when the number of vertices $N$ increases from 1000 to 2000 -- see Table \ref{tab:net}. All in all, only the degree, the strength and the accessibility (Fig. \ref{fig:acc}) of each vertex proved useful on prediction of its spreading potential, considering all the network topologies so far discussed.

\begin{figure}[!ht]
\centering
\includegraphics[width=\columnwidth]{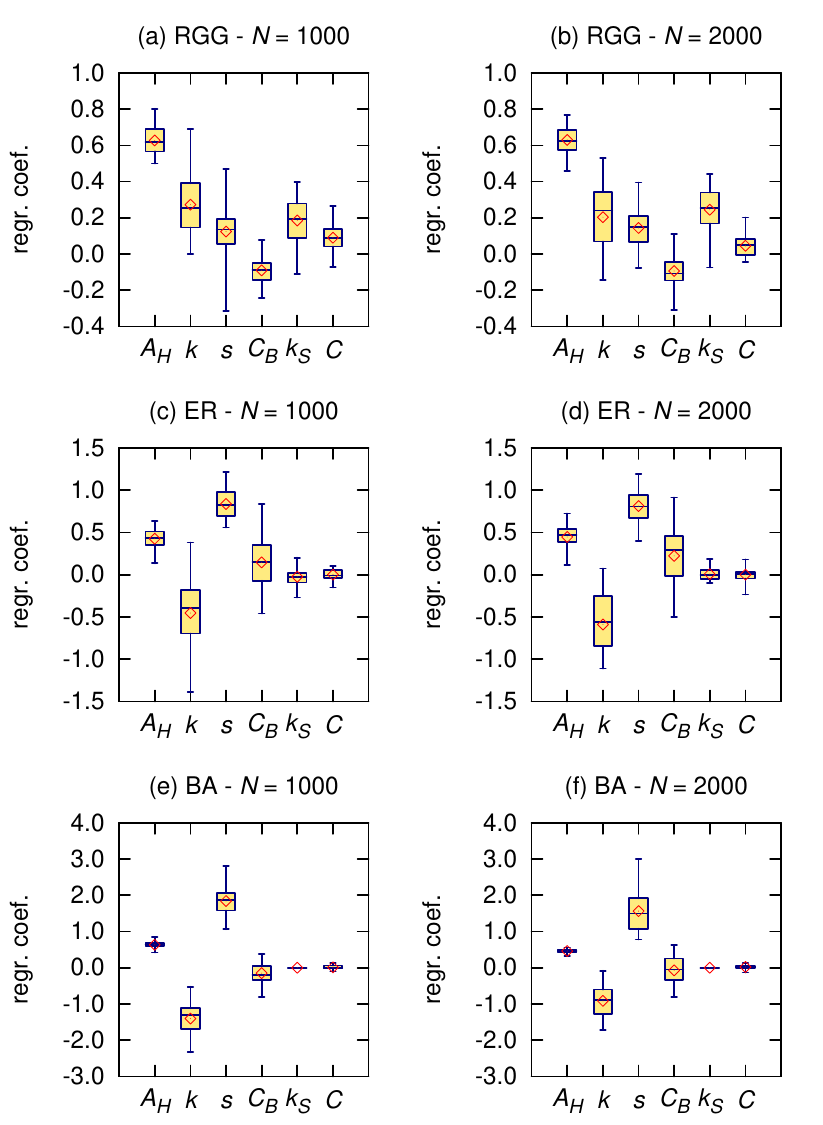}
\caption{\label{fig:regression}Regression coefficients between the epidemic total prevalence and the accessibility $A_H$, the degree $k$, the strength $s$, the betweenness centrality $C_B$, the $k$-shell index $k_S$ and the clustering coefficient $C$, across all the samples for each network class. }
\end{figure}

A natural consequence of the high correlation found between the accessibility $A_H$, the degree $k$ and the strength $s$ of the vertices and their respective spreading potential is that such measures exhibit high correlation among themselves. Thus we also observed the epidemic outbreak size to be a function of the whole set of measures, through a linear regression approach (e.g. \cite{draper_1998}), in an attempt to identify which of the vertex characteristics are more related to its spreading potential. Here we assume that the epidemic outbreak size $r_\infty$ is a linear combination of the six measures considered, and identify the coefficients for each case. Fig. \ref{fig:regression} illustrates the regression analysis, in terms of the {\it regression coefficients} (i.e. the coefficients mentioned beforehand) found for each measure, where we observe that the spreading potential of the vertices is more related to their respective strength $s$, considering the non-geographic models, whereas the accessibility $A_H$ and to a lower level the degree $k$ better explain the spreading potential in the case of RGG model. 

\subsection{A real-world case: US Air Transportation Network}

\begin{figure}[th]
\centering
\includegraphics[width=.9\columnwidth]{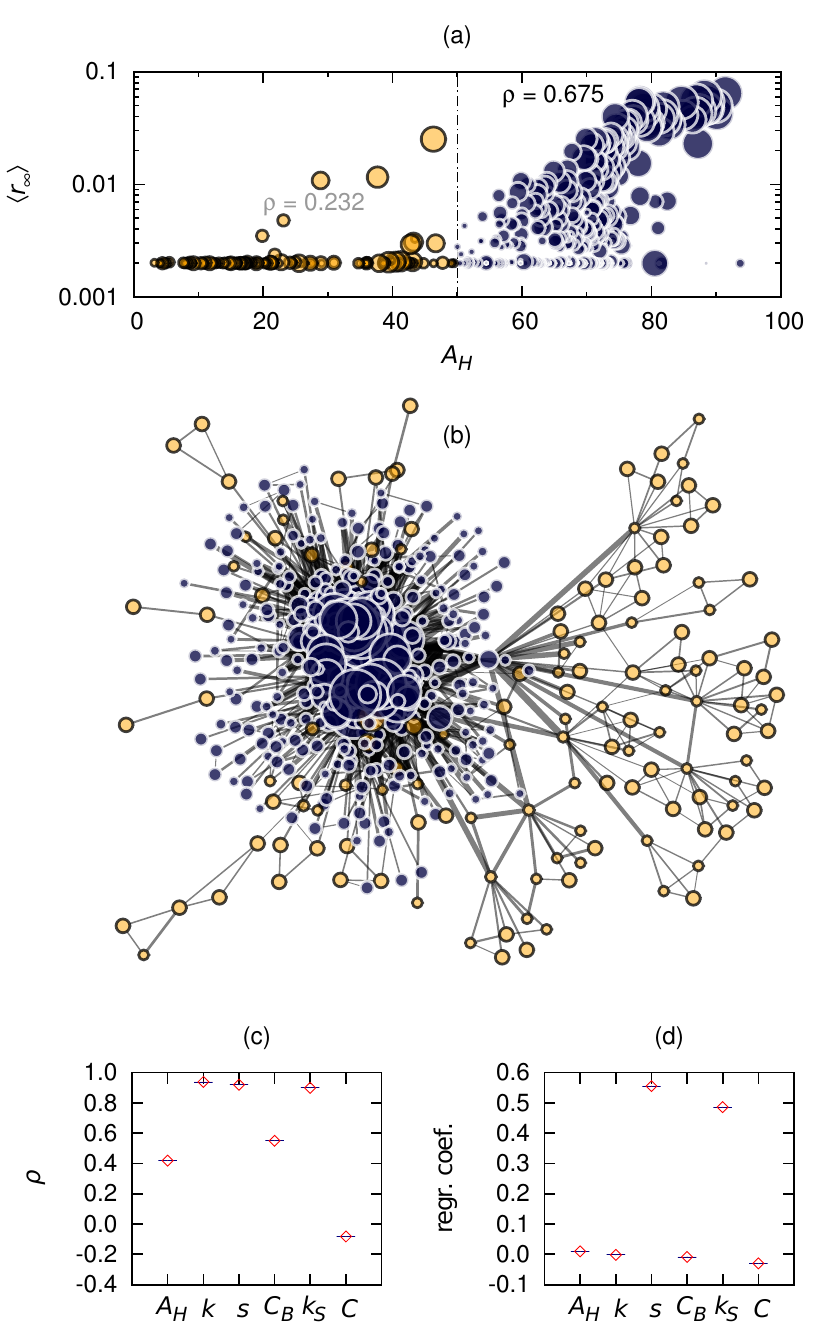}
\caption{\label{fig:usair}US Air Transportation Network: (a) Average outbreak size as function of the vertex accessibility, considering all the 500 vertices. Circles radii proportional to the degrees of corresponding vertices, separated into 2 subsets according to the accessibility: lower ($A_H\leq 50$) and higher ($A_H>50$) values, represented with light and dark tones, respectively. Pearson correlation coefficient is given for each group separately. (b) The respective graph. Thickness of connections is related to the passenger seat availability (proportional to the transmission rate among the sites). Vertex size is proportional to the epidemic outbreak size. The emergence of a ``core'' with high epidemic outbreak is evident, almost exclusively formed by most accessible vertices. (c) Pearson correlation coefficient $\rho$ between the average outbreak size and the accessibility $A_H$ (considering vertices with high accessibility), the degree $k$, the strength $s$, the betweenness centrality $C_B$, 
the $k$-shell index $k_S$ and the clustering coefficient $C$. (d) Regression coefficients for such measures, considering linear regression.}
\end{figure}

We extend the analysis by the incorporation of a ``real-world'' network on the study -- the US Air Transportation Network \cite{colizza_2007}. Such system corresponds to a weighted undirected network whose vertices represent the 500 busiest airports in the United States, according to available data, and weighted connections indicate the existence of flights connecting two airports, the weight corresponding to the number of available passenger seats for the period of one year. The hypothesis here adopted is that the transmission rate is directly proportional to seat availability, thus we consider the spreading rate between airports $i$ and $j$ as $\beta_{ij}=S(i,j)/\max_{i,j}[S(i,j)]$, where $S(i,j)$ is the yearly seat availability between such points. The epidemic threshold for an equivalent random network under homogeneous spreading regime and with the average spreading rate $\langle \beta_{ij}\rangle$ obtained for the airport network, is $\lambda_c\approx 0.019$ -- as estimated through Eq. (\ref{eq:thresh}). In this text we take $\lambda\approx 0.0214$, by fixing the recovery rate at $\mu=3.18$. We explore the behavior of the epidemic spread by considering all the 500 airports as source, and average the epidemic outbreak size starting at each point over 100 realizations of the SIR model, identical to the analysis performed for the modeled networks. High correlation is observed between the epidemic outbreak size and vertices' degrees, strengths and $k$-shell indices -- 0.94, 0.92 and 0.90, respectively -- and a moderate correlation for the accessibility, $\rho({\bf r_\infty},{\bf A_H})=0.42$ (see Fig. \ref{fig:usair}). Regarding the latter measurement, we verify that low correlation is consequence of behavior observed for vertices with low accessibility --  $A_H(i)\leq 50$. With a few exceptions, it has been observed that epidemic processes started at such vertices are not capable to enter endemic phase -- see Fig. \ref{fig:usair}(a). In the case of airports with higher accessibility, on the other hand, the behavior is to a large extent similar to that observed in the case of modeled networks, and the Pearson correlation coefficient increases to $\rho({\bf r_\infty},{\bf A_H})=0.675$. Furthermore, we see from Fig. \ref{fig:usair}(b) that the latter set of vertices define a ``core'' in the network --  - with the remaining vertices occupying ``the border'' of the corresponding graph. The division of the vertices into border and non-border sets by means of accessibility has also been observed \cite{travencolo_2008}, allowing, for example, one to explore topological and dynamical aspects of the network free of border effects \cite{viana_2010}. Such division is made in terms of a threshold accessibility value, such that vertices with accessibility below the threshold are taken as belonging to the border region. For the current case of study, such threshold is given in terms of the spreading dynamics. The increase of epidemic process effectiveness -- e.g. by reduction of the epidemic recovery rate $\mu$ -- is expected to reduce the accessibility threshold, since a largest fraction of vertices is expected to be effective in spreading the disease, and vice versa. Plot (d) gives the respective linear regression analysis between the spreading potential of each vertex and its respective topological attributes. It can be seen for this particular network that the strength $s$ and the $k$-shell index $k_S$ of each point mostly determine its spreading potential.

\section{Conclusions}

In this text the behavior of heterogeneous spreading processes throughout networked systems has been explored in terms of diverse attributes of the originating vertices, including the degree and centrality measurements, such as the betweenness centrality,  the $k$-shell index and the accessibility, a novel measure related to the communicability of the network. A geographic model has been considered, the RGG, as well as ER and BA networks. We also investigated possible correlations between the spreading potential and individual attributes of the source vertices for a real-world network, the US Air Transportation Network. 

For the modeled structures a particular distribution of the transmission rate across the connections was derived and adopted, promoting the heterogeneity of epidemic process. For the airport network, on the other hand, the distribution is empirically given as function of the amount of available passenger seats between the airports. Among the vertex attributes used in the text, only the degree, the strength (weighted degree) and the accessibility showed good correlation with the average epidemic prevalence, considering all the topologies studied, in contrast with what was observed for the local connectivity, estimated by the clustering coefficient, which cannot be taken into account on the prediction of the spreading potential of the vertex. The $k$-shell index showed high correlation with the epidemic prevalence in all the cases, except BA networks, where all the vertices were assigned the same $k$-shell index, which makes its employment as prediction tool for this particular case inadequate, in opposition to what was observed in small-world systems \cite{kitsak_2010}. For the RGG model, where spatial limitations are imposed to the connections, betweenness centrality lacks correlation with the epidemic prevalence. A moderate correlation for the betweenness centrality is observed in the airport network, and higher correlation is achieved in the case of ER and BA models. Finally, owing to the particular topology of the airport network, spreading processes starting at vertices with lower accessibility are unable to persist and enter the endemic phase, with exceptions on a few airports, verified to receive more connections than the remaining. If we consider the airports with accessibility higher than a determinate value, on the other hand, the epidemic processes persist, with average prevalence proportional to the accessibility, resulting in higher correlation. Our results demonstrate the influence of topological aspects of the network as a whole over the prediction of the spreading potential at the individual level. Such influence is observed on the modification or to a large extent on the elimination of the correlation between vertex characteristics and dynamic properties of the epidemic process. We believe from such results that the prediction of potential spreaders in networks, specifically without {\it a priori} knowledge about the overall system topology, should consider a combination of individual features, rather than be based only on a single attribute, which can be misleading and suggest inefficient eradication policies. Possible further investigations include observing how the correlations evolve with the epidemic process -- it is plausible, for example, to expect higher correlation between the clustering coefficient of the source vertex and the number of infected individuals at early stages of the spread, since at this stage only vertices at the proximity of the source are likely to be contaminated.

\begin{acknowledgments}
R. A. P. Silva and M. P. Viana thank FAPESP for financial support (grants 2007/54742-7 and 2010/16310-0, respectively). L. da F. Costa also thanks FAPESP (2005/00587-5) and CNPq (301303/2006-1 and 573583/2008-0) for sponsorship.
\end{acknowledgments}

\bibliographystyle{unsrt}
\bibliography{acc}

\begin{thebibliography}{10}

\bibitem{costa_et_al_2011_appli}
L.~{\mbox da F}. Costa, O.~N. {Oliveira Jr.}, G.~Travieso, F.~A. Rodrigues,
  P.~R. Villas~Boas, L.~Antiqueira, M.~P. Viana, and L.~E.~C. Rocha.
\newblock Analyzing and modeling real-world phenomena with complex networks: a
  survey of applications.
\newblock {\em Adv. Phys.}, 60(3):329--412, 5 2011.

\bibitem{danon_2011}
L.~Danon, A.~P. Ford, T.~House, C.~P. Jewell, M.~J. Keeling, G.~O. Roberts,
  J.~V. Ross, and M.~C. Vernon.
\newblock Networks and the epidemiology of infectious disease.
\newblock {\em Interdisc. Persp. Infect. Diseases}, 2011:284909, 2011.

\bibitem{barabasi_1999}
A.-L. Barab{\'a}si and R.~Albert.
\newblock Emergence of scaling in random networks.
\newblock {\em Science}, 286(5439):509--512, 1999.

\bibitem{pastor-satorras_2001}
R.~Pastor-Satorras and A.~Vespignani.
\newblock Epidemic spreading in scale-free networks.
\newblock {\em Phys. Rev. Lett.}, 86(14):3200--3203, 2001.

\bibitem{dezso_2002}
Z.~Dezs{\"o} and A.-L. Barab{\'a}si.
\newblock Halting viruses in scale-free networks.
\newblock {\em Phys. Rev. E}, 65(5):055103(R), 2002.

\bibitem{moore_2000}
C.~Moore and M.~E.~J. Newman.
\newblock Epidemics and percolation in small-world networks.
\newblock {\em Phys. Rev. E}, 61(5):5678--5682, 2000.

\bibitem{kitsak_2010}
M.~Kitsak, L.~K. Gallos, S.~Havlin, F.~Liljeros, L.~Muchnik, H.~E. Stanley, and
  H.~A. Makse.
\newblock Identification of influential spreaders in complex networks.
\newblock {\em Nature Phys.}, 6(11):888--893, 2010.

\bibitem{carmi_2007}
S.~Carmi, S.~Havlin, S.~Kirkpatrick, Y.~Shavitt, and E.~Shir.
\newblock A model of internet topology using $k$-shell decomposition.
\newblock {\em Proc. Nat. Acad. Sci. USA}, 104(27):11150--11154, 2007.

\bibitem{gang_2005}
Y.~Gang, Z.~Tao, W.~Jie, F.~Zhong-Qian, and W.~Bing-Hong.
\newblock Epidemic spread in weighted scale-free networks.
\newblock {\em Chinese Phys. Lett.}, 22(2):510, 2005.

\bibitem{schumm_2007}
P.~Schumm, C.~Scoglio, D.~Gruenbacher, and T.~Easton.
\newblock Epidemic spreading on weighted contact networks.
\newblock In {\em Proc. BIONETICS'07}, pages 201--208, 2007.

\bibitem{britton_2011}
T.~Britton, M.~Deijfen, and F.~Liljeros.
\newblock A weighted configuration model and inhomogeneous epidemics.
\newblock {\em J. Stat. Phys.}, 145:1368--1384, 2011.

\bibitem{kao_2006}
R.~R. Kao, L.~Danon, D.~M. Green, and I.~Z. Kiss.
\newblock Demographic structure and pathogen dynamics on the network of
  livestock movements in {Great Britain}.
\newblock {\em Proc. R. Soc. B}, 273(1597):1999--2007, 2006.

\bibitem{colizza_2007}
V.~Colizza, R.~Pastor-Satorras, and A.~Vespignani.
\newblock Reaction--diffusion processes and metapopulation models in
  heterogeneous networks.
\newblock {\em Nature Phys.}, 3(4):276--282, mar 2007.

\bibitem{molloy_1995}
M.~Molloy and B.~Reed.
\newblock A critical point for random graphs with a given degree sequence.
\newblock {\em Rand. Struct. Alg.}, 6(2--3):161--180, 1995.

\bibitem{diekmann_2000}
O.~Diekmann and J.~A.~P. Heesterbeek.
\newblock {\em Mathematical Epidemiology of Infectious Diseases -- Model
  Building, Analysis and Interpretation}.
\newblock Oxford University Press, Oxford, 1999.

\bibitem{keeling_2008}
Matt~J. Keeling and Pejman Rohani.
\newblock {\em Modeling infectious diseases in humans and animals}.
\newblock Princeton University Press, Princeton, 2008.

\bibitem{travencolo_2008}
B.~A.~N. Traven\c{c}olo and L.~{\mbox da F}. Costa.
\newblock Accessibility in complex networks.
\newblock {\em Phys. Lett. A}, 373(1):89--95, oct 2008.

\bibitem{halbert_2004}
S.~E. Halbert and K.~L. Manjunath.
\newblock Asian citrus psyllids (sternorrhyncha : Psyllidae) and greening
  disease of citrus: A literature review and assessment of risk in florida.
\newblock {\em Florida Entom.}, 87(3):330--353, 2004.

\bibitem{dall_2002}
J.~Dall and M.~Christensen.
\newblock Random geometric graphs.
\newblock {\em Phys. Rev. E}, 66(1):016121, 2002.

\bibitem{barthelemy_2011}
M.~Barth{\'e}lemy.
\newblock Spatial networks.
\newblock {\em Phys. Rep.}, 499(1--3):1--101, 2011.

\bibitem{Note1}
If edges are defined with probability $p$, then on a graph with $N$ vertices
  the expected number of edges is $pN(N-1)/2$. In other words, the average
  degree of the graph is $p(N-1)$.

\bibitem{erdos_1959}
P.~Erd{\"o}s and A.~R{\'e}nyi.
\newblock On random graphs, i.
\newblock {\em Publi. Mathematicae}, 6:290--297, 1959.

\bibitem{costa_et_al_2007_meas}
L.~{\mbox da F}. Costa, F.~A. Rodrigues, G.~Travieso, and P.~R. Villas~Boas.
\newblock Characterization of complex networks: A survey of measurements.
\newblock {\em Adv. Phys.}, 56(1):167--242, 2007.

\bibitem{freeman_1977}
L.~C. Freeman.
\newblock A set of measures of centrality based on betweenness.
\newblock {\em Sociometry}, 40(1):35--41, March 1977.

\bibitem{barrat_2008}
A.~Barrat, M.~Barth{\'e}lemy, and A.~Vespignani.
\newblock {\em Dynamical Processes on Complex Networks}.
\newblock Cambridge University Press, Cambridge, UK, 2008.

\bibitem{draper_1998}
N.~R. Draper and H.~Smith.
\newblock {\em Applied Regression Analysis}.
\newblock Wiley Series in Probability and Statistics. John Wiley \& Sons, New
  York, NY, USA, 3 edition, 1998.

\bibitem{viana_2010}
M.~P. Viana, B.~A.~N. Traven\c{c}olo, E.~Tanck, and L.~{\mbox da F}. Costa.
\newblock Characterizing topological and dynamical properties of complex
  networks without border effects.
\newblock {\em Physica A}, 389(8):1771--1778, 2010.

\end{thebibliography}

\end{document}